\newcommand{\CLASSINPUTinnersidemargin}{16.9mm}
\newcommand{\CLASSINPUTtoptextmargin}{19mm}
\newcommand{\CLASSINPUTbottomtextmargin}{25mm}
\documentclass[conference,a4paper]{IEEEtran}
\ifCLASSINFOpdf
   \usepackage[pdftex]{graphicx}
\else
\fi
\usepackage[cmex10]{amsmath}
\usepackage{xcolor}
\usepackage{comment}
\usepackage{multirow}
\usepackage{array}
\begin{document}
%
\title{Analysis of the Implication of Current Limits in Grid Forming Wind Farm}

\author{\IEEEauthorblockN{Kanakesh Vatta Kkuni\\ Guangya Yang}
\IEEEauthorblockA{\textit{Department of Electrical Engineering}\\\textit{Technical University of Denmark}\\
\textit{2800 Kgs. Lyngby, Denmark}\\
Email: \textit{kvkhu@elektro.dtu.dk}}
\and
\IEEEauthorblockN{Thyge Knüppel}
\IEEEauthorblockA{\textit{Siemens Gamesa}}
Email: \textit{Thyge.Knueppel@siemensgamesa.com}}


%


\maketitle

\begin{abstract}
There is an ongoing trend of reduction in short circuit power at the grid connection point due to decommissioning of synchronous generation plants causing system strength issues in wind power plants. Whereas wind power plant rating and export cable length are increasing, further weakening the system strength and accompanied by stability challenges. Under such a scenario, a grid forming control demonstrated to operate in a weaker system has value creation potential for application in wind turbine generators. In addition, the grid forming control can also enable a wind power plant to operate in islanded mode, provide inertially and phase jump active power support. However, the application of grid forming control has challenges because grid forming control applied to a power converter (GFC) has a voltage source behavior and does not stiffly control the grid side active power and thus requires a separate current limiting mechanism. However, there could be potential challenges in maintaining the synchronism of GFC when the current limit is triggered, particularly during the grid voltage phase jump event. Modeling and capturing such a phenomenon is a challenge in a wind farm with many wind turbines. To that end, this paper investigates the modeling adequacy of the aggregated GFC-WF to a single GF-WTG of total WF rating in capturing GFC-WF dynamics. The challenges related to loss of synchronization stability when one or more wind turbine generators enter current limited operation during a grid phase jump events are also evaluated in this paper. 
\end{abstract}


%
\IEEEpeerreviewmaketitle

\section{Introduction}
The high penetration of power converter-enabled renewable energy (RE) generation and a reduction in the share of synchronous machines have introduced several challenges to the power system. Some of the expected challenges with a sizeable nonsynchronous-based power infeed are reduced levels of inertia and short circuit power, which may, ultimately, lead to adverse control interactions in the system and a need to manage the penetration level of nonsynchronous-based power infeed if the control is not sufficiently robust to the weakening of the system. For instance, past studies on the GB  transmission system model showed that it is impossible to increase the nonsynchronous renewable penetration above 65$\%$ \cite{urdal2015system}.

\begin{table}[b]
\hrule
Figures and presented values in the paper should not be used to judge the performance of any Siemens Gamesa Renewable Energy wind turbines as they are solely presented for demonstration purpose. 

Any opinions or analysis contained in this paper are the opinions of the authors and not necessarily the same as those of Siemens Gamesa Renewable Energy A/S. 

Nothing in this paper shall be construed as acknowledgement of feature implementation in offered Siemens Gamesa Renewable Energy wind turbines.
\end{table}

The grid forming control, which enables a near voltage source behavior for the implemented power converter, is one of the potential solutions to mitigate the challenges caused to high nonsynchronous generation \cite{ENTSO-E2019, NGSOF}. To that end, the application of grid forming control on WTG in offshore wind farms is expected to  facilitate larger power transfer through the HVAC export cables without stability constraints thus reducing the cost. The application of grid forming control on type IV grid forming wind turbines can also facilitate black start, and islanding operation. A small-scale field trial of wind turbine generators (WTGs) in grid forming mode demonstrating the islanded mode and different levels of inertia contribution is reported in \cite{brogan2018experience}. And a more extensive field trial of a wind farm with 23 WTG of each rated 3 MW in grid forming mode is reported in \cite{Roscoe2020} with a focus on the impact of damping.

In recent times, system operators attempt to define the high-level specifications for grid forming converters (GFC). For instance, The National Grid ESO has already published a draft grid code for grid forming convertors' response requirements, and specifications  \cite{NG_GFC}. These specifications mandate that the GFC have a voltage source behavior and provide a near-instantaneous phase jump power and fault current without any additional control loop similar to a synchronous machine. The near-instantaneous current response from the grid forming power converter to a change of grid voltage magnitude or phase is inversely proportional to the impedance between the internal voltage of the GFC and the grid. Therefore, depending on the pre-existing loading of the converter, even a slight phase shift of a few degrees or a small voltage drop can trigger overcurrent. The GFC commonly derives the synchronization from the measured output power \cite{Qoria2020b,zhang2009power}. Therefore, when the overcurrent limiting engages and, effectively, breaks this loop, can result in the loss of synchronization of the GFC. Multiple recent studies have presented the assessment of the synchronizing capability of the grid forming converters under transient events \cite{Qoria2020b,Wu2019}. It has been identified that regardless of the type of current limiting algorithm employed, the stability margin for maintaining the synchronization of the grid forming converter drops significantly when the grid forming converter enters the current limited operation.

Such potential instabilities present a modeling and analysis challenge in assessing grid forming wind farm (GFC-WF), which consisting of multiple WTGs implemented with grid forming control.  When one or more GFC-WTG  in WF enter the current limited operation during phase jump events, which happens almost instantaneously, the aggregated modeling of the WF could potentially fail to capture such instabilities. To that end, this paper focuses the analysis and studies on the following questions
\begin{itemize}
    \item How adequate is aggregating the GFC-WF to a single GF-WTG of total WF rating in capturing GFC-WF dynamics?
    \item What are the challenges incurred when one or several wind turbines in a wind park enter the current limited operation during a grid phase jump event?
\end{itemize}


 
\section{System and Modelling Description}
\begin{figure*}
  \includegraphics[width=0.9\textwidth,height=10cm]{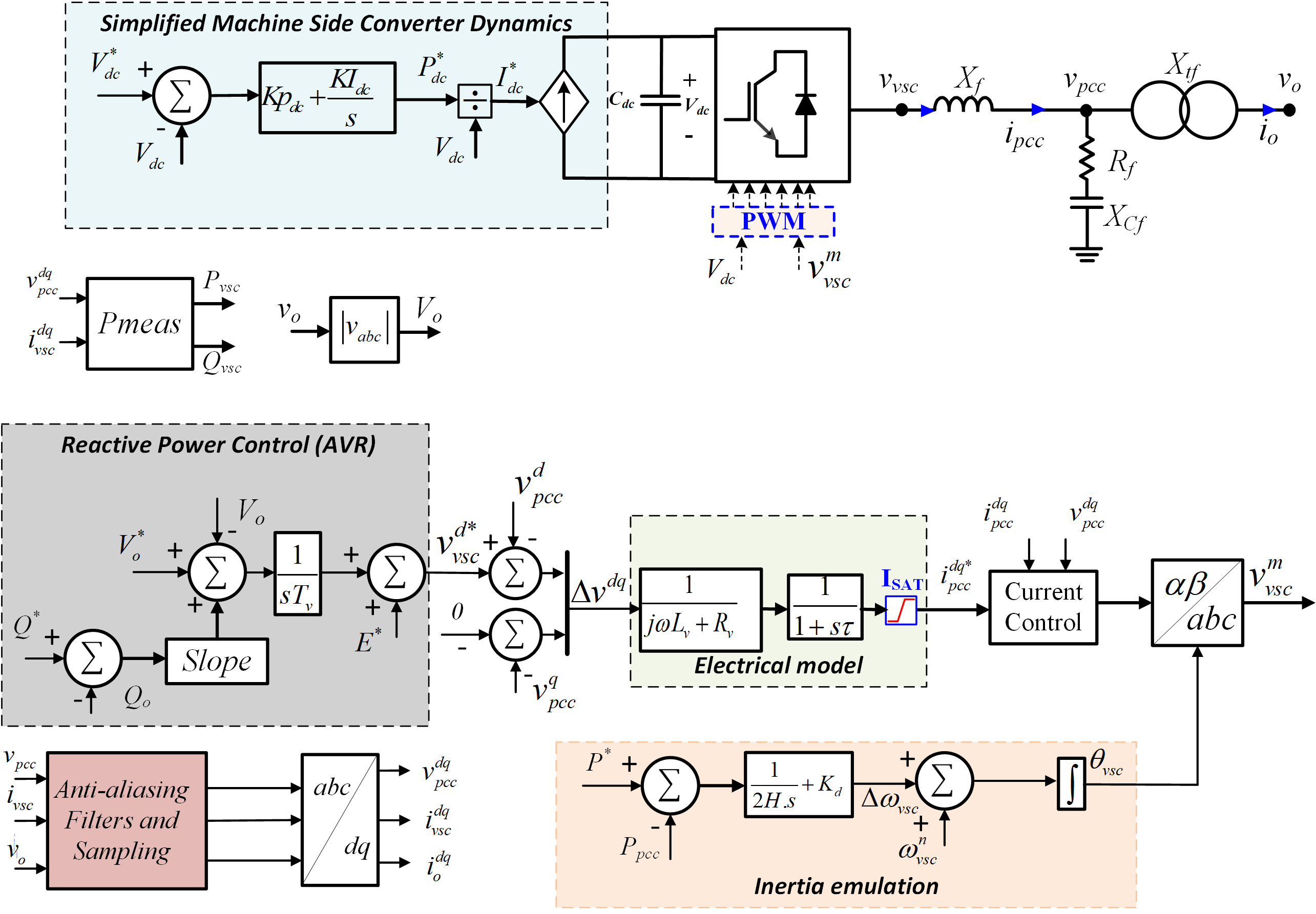}
  \caption{Converter and the implemented grid forming control}
  \label{fig:struct}
\end{figure*}

The grid forming control is implemented on the wind turbines of the benchmark 420 MW wind farm (WF) model developed by the CIGRE C4.49 working group \cite{cigre1}. In this section the hardware and control implemented on the WTG is described. In addition, the  modelling methodology to develop the EMT WF model aggregated at the WF level and string level are presented.    
\subsection{Grid forming WTG model}
The grid forming control has several realizations reported in the literature. 
The major differences between reported topologies are how the inertial (virtual rotor) or active power control is implemented\cite{chen2020modelling}. In addition, the difference between the reported GFC can also be classified based on the presence of inner control loops such as current and voltage control \cite{kkuni2021comparative}. A GFC with an inner current control is chosen in this study \cite{zhang2016synchronous} for the sake of more accessible and faster current limit implementation. 

The structure of the control implemented and the single line diagram of the converter is shown in Fig. \ref{fig:struct}. Similar to a synchronous machine model, the GFC control can be partitioned to rotor dynamics (inertia emulation), electrical equivalent model, and excitation system model (reactive power control). 
 When selecting the inertial characteristics for the GFC, the limitations of the converter hardware and the sourcing of energy for the response must be considered. If an underdamped response, as in the case of a synchronous machine, is required, the peak response of both current and power must be considered when evaluating hardware capability. Furthermore, it is challenging to increase the damping of the GFC when a swing equation-based inertial emulation is implemented. However, with a proportional-integral-based inertial emulation \cite{zhang2015comparison,taul2019current} implemented in this study, the damping of the GFC response can be increased by increasing the proportional gain constant $K_d$, and the integral gain is equal to $\frac{1}{2H}$ where H is the inertia constant in seconds. 
 
 A magnitude limiter for the current reference is implemented for the current limiting, the current magnitude is limited to 1.2 pu. The current vector $i_{pcc}^{dqLi{m^*}}$ in current limited mode is given by 
 \begin{equation}
i_{pcc}^{dqLi{m^*}}=\frac{1}{KC_{lim}}*I^{dq*}_{pcc},\text{ where, }   KC_{lim}= \frac{\left| I^{dq*}_{pcc} \right|}{1.2}
\label{eqn:Ilim}
\end{equation}

$\left| I^{dq*}_{pcc} \right|$ is the unsaturated reference current vector magnitude,  thus the vector $i^{dqLim^*}_{pcc}$ is of the magnitude 1.2 pu during current limited operation. During the overcurrent limiting under the grid frequency or phase event, the measured power becomes insensitive to the synchronization/inertial loop output and can potentially result in the loss of synchronization of the GFC.

The excitation system consists of a voltage controller that maintains the medium voltage bus (66 kV) at reference voltage with a reactive power slope. The electrical model consists primarily of an emulated reactor with an inductance of $L_v$ and resistance of $R_v$, a first-order low pass filter, and a DQ decoupled current control. The electrical model is realized in the reference frame defined by the virtual rotor. The difference between the voltages of the internal voltage source ($v_{vsc}$) and ac capacitor terminal ac voltage ($v_{pcc}$) is applied to the admittance of the virtual reactor, which generates the current references for the current control. The d-q superscript for the variables implies the variables are the of the direct and quadrature components of the variables depicted in the synchronously rotating reference frame defined by the virtual rotor of the GFC ($\theta_{vsc}$).

\begin{table}[]
\caption{Parameters of WTG converter system \cite{cigre1}}
\begin{tabular}{|l|l|l|}
\hline
 \textbf{Name} & \textbf{Value} &  \textbf{Description [unit]} \\ \hline
 Sbase & 12  & Rated Power [MW] \\ \hline
 fsw & 2950 &Switching frequency [Hz] \\ \hline
 fsamp & 2fsw &Sampling frequency [Hz] \\ \hline
 kmod & $\sqrt{3}$/2 &Modulation constant (sine PWM) [pu] \\ \hline
 vdcnom & 2 &Nominal dc voltage [pu] = 1.38 kV \\ \hline
 rf & lf/20 &Filter resistance [pu] \\ \hline
 lf & 0.1055776 &Filter inductance inverter side [pu] \\ \hline
 rcf & 0.003& Filter resistance [pu] \\ \hline
 cf & 0.0757204 &Filter capacitance [pu] \\ \hline
 cdc & 6.6654 E10-3& DC capacitor [pu] \\ \hline
 St & 14 &Transformer rating [MVA] \\ \hline
 rt & 0.0054 &Transfomer resistance [pu] \\ \hline
 lt &0.1 &Transfomer inductance [pu] \\ \hline
\end{tabular}
\end{table}

The WTG converter model is depicted in Fig. \ref{fig:struct}. The electrical parameters of the converter are the same as the WTG converter system of the CIGRE benchmark model. It has to be noted that the machine side converter is responsible for maintaining the dc-link voltage, and the grid side converter maintains the power to the ac grid. The machine side power dynamics are neglected and represented by a current source feeding the dc capacitance to simplify the modeling and reduce the computational burden.

 It has to be noted that the retrievable energy stored in dc capacitance for WTG is quite low. For instance, the total equivalent inertial constant (H) that the considered WTG can emulate, by only accounting for the energy stored in the dc-link capacitor is approximately 13 ms. Furthermore, considering the voltage limitations on the dc-link, the actual equivalent inertial constant will be a fraction of that value. Implementing grid forming control with inertia implies not tightly controlling the power during the phase jump, and frequency event as necessitated by emerging specifications of the grid forming control \cite{NG_GFC}.  Hence, inertia and phase jump power needs to be also derived from the machine side. 

The focus of this study is only grid-side converter dynamics in grid forming mode. The following factors need to be added to the modeling to capture the full dc link and machine dynamics for the full-scale evaluation of the GFC-WTG.
\begin{itemize}
    \item The machine model, including the turbine dynamic power ramping capabilities, need to be carefully considered and modeled in detail to study the impact of extreme grid events on the dc link and the machine.
    \item For this study, it is assumed that there is enough generation headroom available, and the WTG is not operating at zero power, in which case incorporating the inertial provision can be challenging.
    \item A dc chopper that clamps the dc-link voltage for avoiding over voltage in the dc side, in practice, the chopper acts to keep the dc-link voltage within a narrow range of its rated value
\end{itemize}

\section{Wind Farm Modelling}

\begin{figure}
  \includegraphics[width=8.5cm]{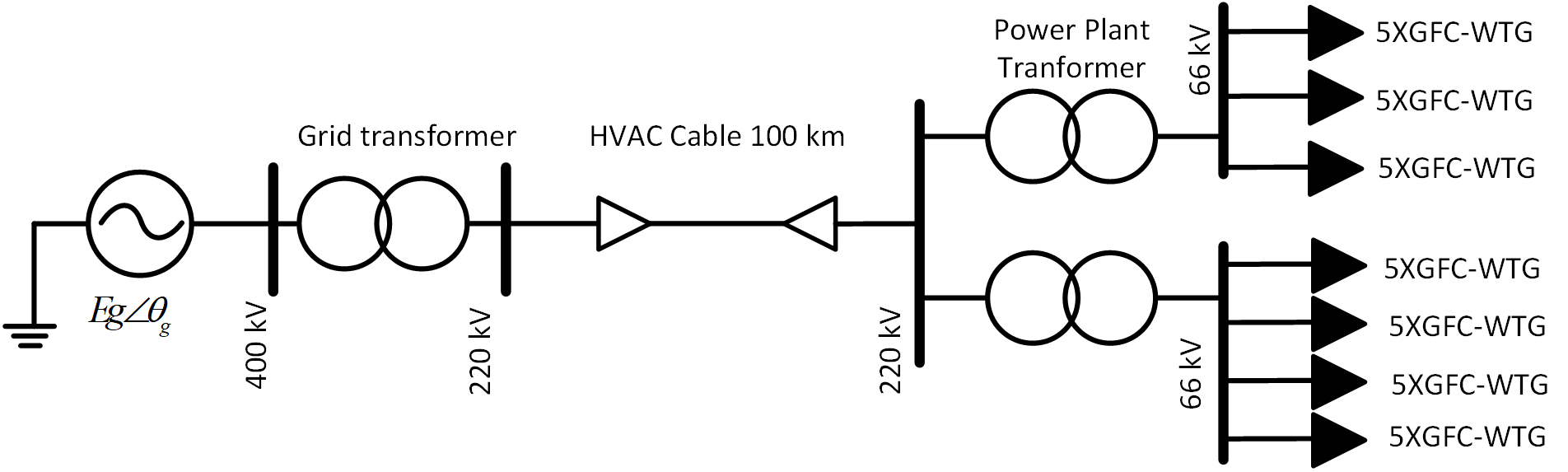}
  \caption{The studied Wind farm layout}
  \label{fig:WF}
\end{figure}

The CIGRE benchmark wind farm consists of 35 WTG of 12 MW each, in this study the control of the WTGs are replaced with grid forming control with current limit functionality. The layout is arranged in 7 strings of 5 WTG each as shown in Fig. \ref{fig:WF}, the WTG's are interconnected by a 66 kV collection cable. The parameters of the collection cables, HVAC cables, and the transformers can be found in \cite{cigre1}. The HVAC cables are modeled as ten interconnected PI sections of inductors and capacitances. And the short circuit MVA at the grid entry point is 3000 MVA at 400 kV.
A detailed EMT simulation of an entire wind farm is computationally intensive, thus aggregating the wind farm is a common practice. In this study, aggregation at two aggregation levels is considered. First, the model is aggregated into a single 420 MW aggregated WTG (FAW). Secondly, the WF is aggregated to string level (5X12 MW WTG). Thus there Seven strings of 60 MW each in the string aggregate wind farm (SAW). In both cases, the collector cables are aggregated using an electrical equivalent model.

The modeling and simulation are conducted in MATLAB/Simulink. Firstly the EMT model of the WF with a fully aggregated single 420 MW GFC- WTG with the fully aggregated collector is implemented using MATLAB/SimPowerSystems components. From a computational perspective, the aggregated model is very effective. Nevertheless, the fully aggregated model is not adequate in capturing accurate dynamics of the WF, especially when the generation of the WTG's and the network between the WTGs and aggregation point is unsymmetric. The aforementioned is often the case due to the wake effect and spatial distribution of the wind farm.
On the other hand, a full EMT model representing every WTG and cable in detail is not practical due to the computational burden. Therefore, a vectorized programming methodology of the WF discussed in \cite{kunjumuhammed2016adequacy} is employed in this study. The following steps are followed to develop and verify the vectorized dq model of the wind farm.

\begin{enumerate}
    \item EMT model of the WF with a fully aggregated single 420 MW GFC- WTG with the fully aggregated collector is implemented using MATLAB/SimPowerSystems components
    \item The dq domain differential equation-based per unit model of the fully aggregated wind farm (FAW) is developed in MATLAB and verified against the MATLAB/SimPowerSystems model.
    \item The dq domain per unit model is vectorized to developed string aggregated wind farm model (SAW). This implies all the electrical and control parameters in a string aggregated model is a vector of size 7X1 
\end{enumerate}

\begin{figure}
  \includegraphics[width=8.5cm]{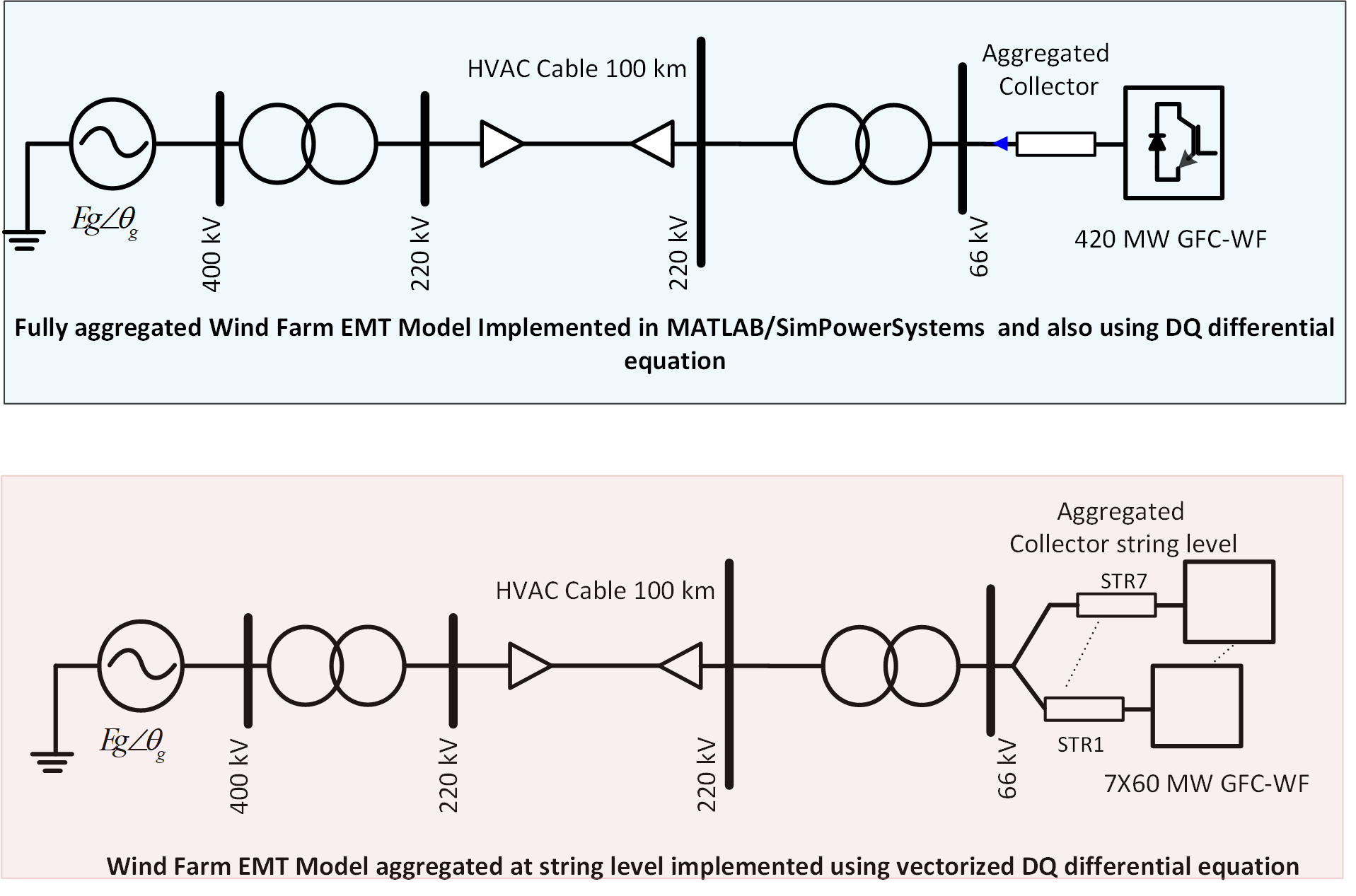}
  \caption{Modelling overview }
  \label{fig:overview}
\end{figure}

The vectorized model facilitates the analysis of WTG strings with heterogeneous parameters. In addition, the dq domain vectorized model is easily scalable to incorporate a WF model with each individual WTG modeled. An overview of the modeling deployed n this study is shown in Fig. \ref{fig:overview}.

\section{Analysis}
The short circuit ratio of the aggregated wind farm at the MV (66 kV) bus is measured to be approximately 2. In addition, the net series reactance between the grid Thevenin voltage source and the PCC point of the WTG is 0.63 pu at 420 MVA base. Thus one can say that the system strength is relatively low for the GFC-WTG operation. Under such system strength, there were challenges in stabilizing this GFC-WTG. The current control bandwidth had to be slowed down for GFC-WTG to operate stably without triggering any instability. However, slowing down the GFC current control bandwidth can contradict GFC requirements which necessitates a near-instantaneous response to grid events. Alternatively, the GFC could be designed without an inner current control loop, which introduces challenges in ensuring a sufficient current limiting action when required. The trade-of in terms of ease/difficulty of the two approaches has not been explored  within this work. Further studies on control design to meet small signal stability and large-signal requirements need to be conducted to maximize the GFC-WTG performance. 
The subsequent subsections present the simulation analysis conducted on the single 420 MW fully aggregated GFC-WTG (FAW) and WF aggregated to 7 strings of 60 MW strings each (or SAW). In addition to discussing the adequacy of the FAW model, the challenges introduced due to triggering the current limit during the grid phase jump events are also presented in the following subsections.

\subsection{Adequacy of the aggregated modelling}

\begin{figure}
    \centering
    \includegraphics[width=\linewidth]{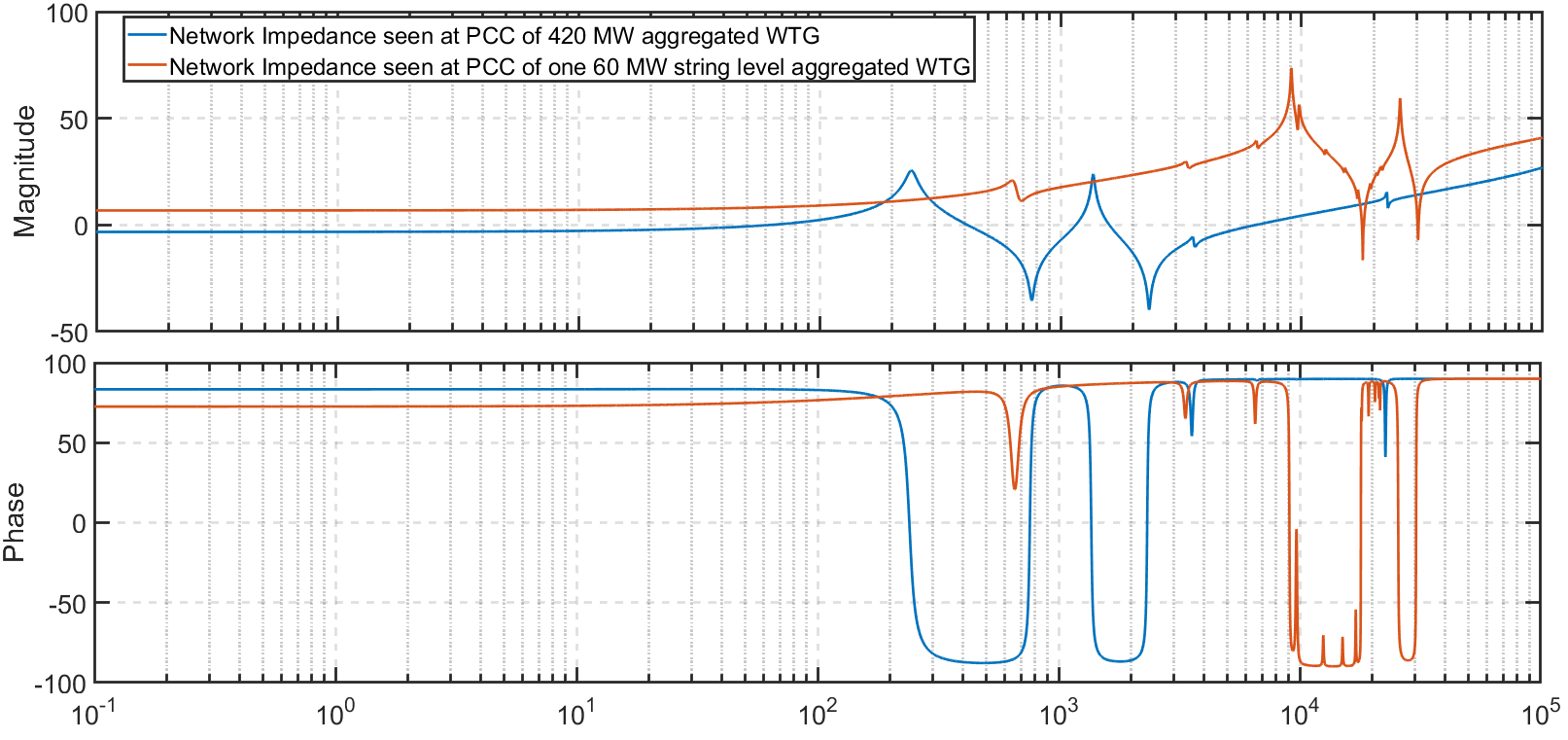}
    \caption{Network Impedances in PU at 420 MVA base power seen by the FAW, and a single string in SAW}
    \label{fig:net_imp}
\end{figure}

\begin{figure}
    \centering
    \includegraphics[width=\linewidth]{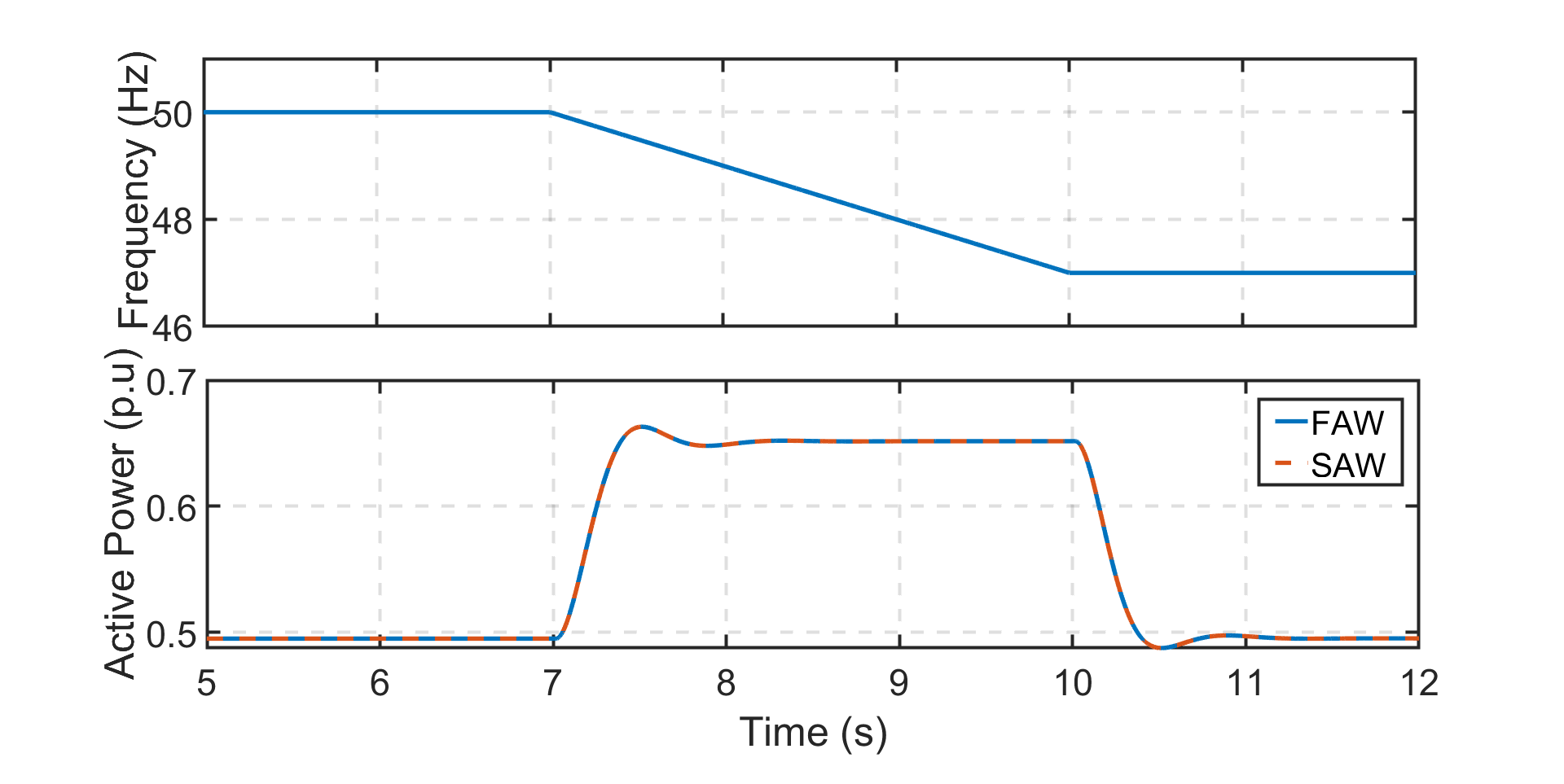}
    \caption{Output power of both SAW and FAW under 1 Hz/s RoCoF}
    \label{fig:rocof}
\end{figure}

It is well known and widely reported in the literature that the aggregated model WF model's dynamic behavior can vary significantly compared to a detailed model considering all the WTGs. For instance, Fig. \ref{fig:net_imp} shows the dynamic network impedance (at 420 MVA base)  seen at the PCC terminal of the fully aggregated WF and the dynamic network impedance (in 60 MVA base) seen by a single string of WTGs, which are significantly different. Thus confirming the dynamics exhibited by the aggregated model can be different from the string level aggregated WF.
In this subsection, all the analysis has been conducted by disabling the current limiter. To begin with, both the FAW and SAW and subjected to a rate of change of frequency (RoCoF) of 1 Hz/s from 50 hz to 47 Hz with programmed inertia of H= 4 s and the damping ratio is designed to be 0.7. Before the RoCoF event, each of the WTGs is generating 0.5 pu of active power. The results of the RoCoF event are shown in Fig. \ref{fig:rocof}. The inertial active power output ($P_{inert}$) matches the programmed inertia corresponding to H= 4s.
\begin{equation}
    P_{inert}=\frac{2H}{Nominal Frequency}=0.16 pu
\end{equation}
It is seen that the dynamics of the power measured at the high voltage side of the power plant transformer remain the same for both models. This equivalency is because all the strings modelled here are symmetrical, i.e, all the electrical parameters of the array cables, transformer, control and generation of all the 7x60 MW GFC-WTG remain the same and are also in parallel. 

\begin{figure}
    \centering
    \includegraphics[width=\linewidth]{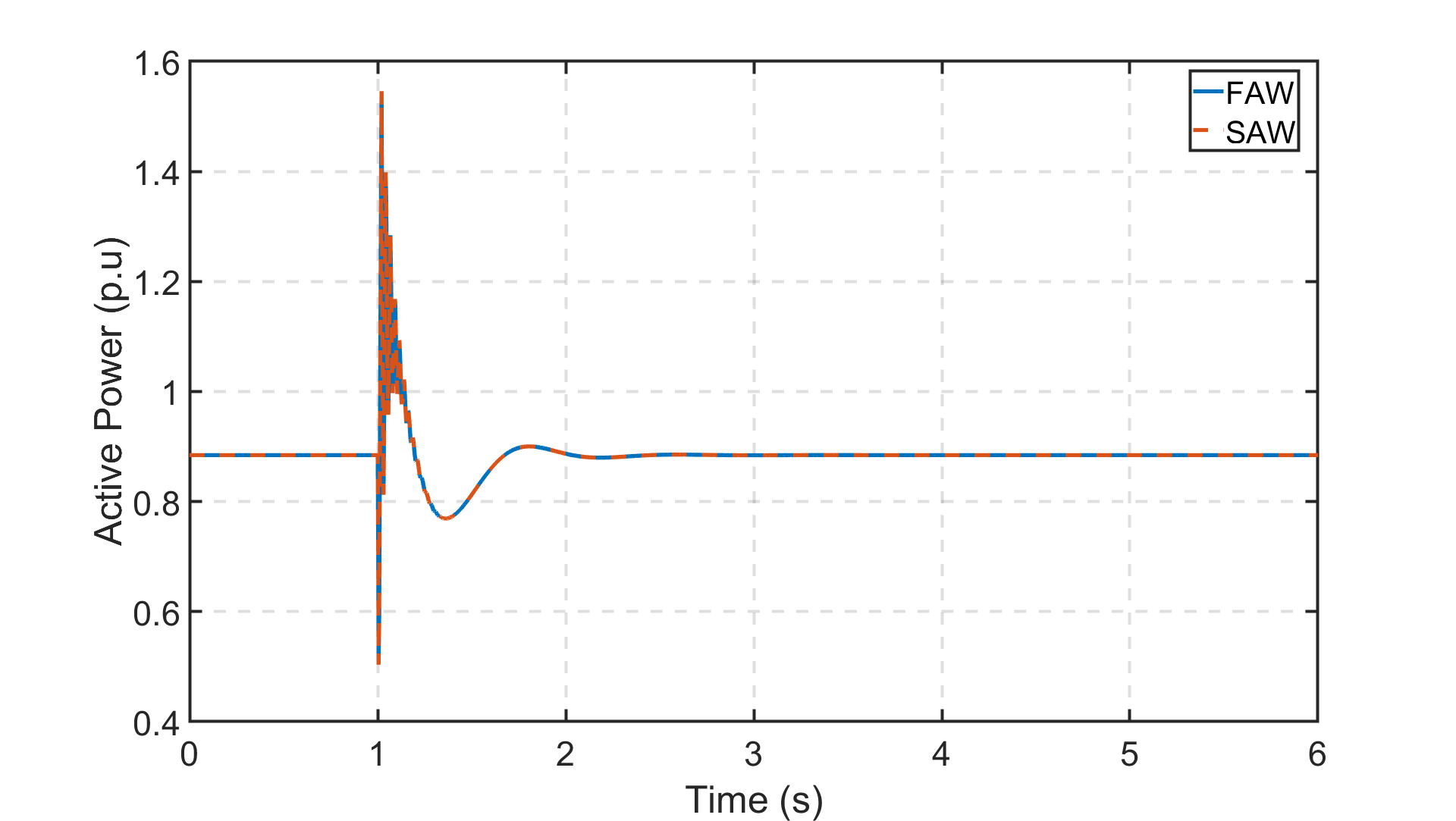}
    \caption{Output power of both SAW and FAW under a $15^{\circ}$ grid phase jump event when each of the 7 strings are at 0.9 pu power generation}
    \label{fig:phae shift 1}
\end{figure}

Similarly, during grid phase jump event when the power generation among the strings are equal, the power ouput from both FAW and SAW remains the same as depicted in Fig. \ref{fig:phae shift 1}. 

\begin{figure}
    \centering
    \includegraphics[width=\linewidth]{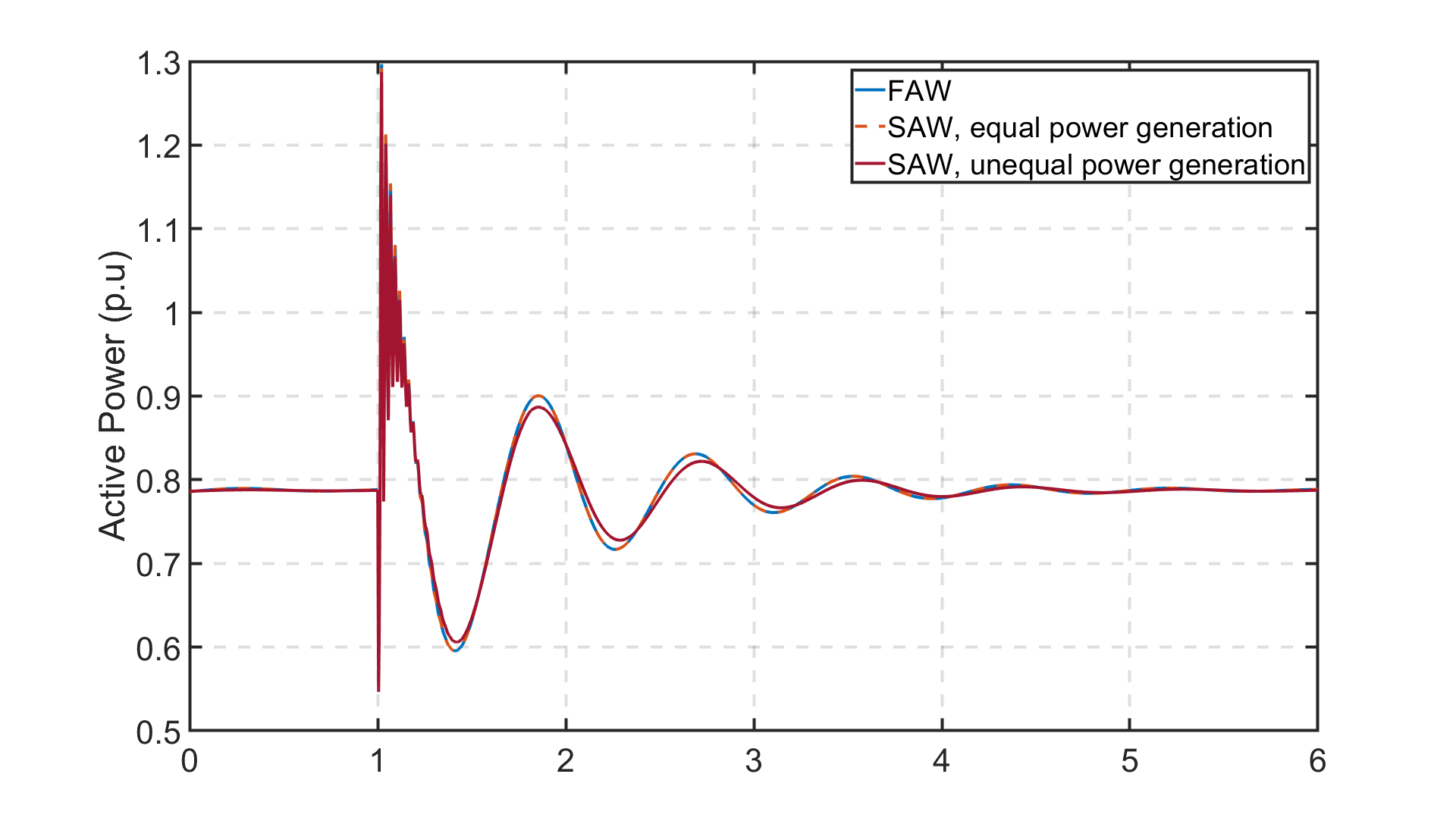}
    \caption{Output power of both SAW and FAW under a $15^{\circ}$ grid phase jump event with equal and unequal generation among the strings}
    \label{fig:phase shift 3}
\end{figure}

\begin{figure}
    \centering
    \includegraphics[width=\linewidth]{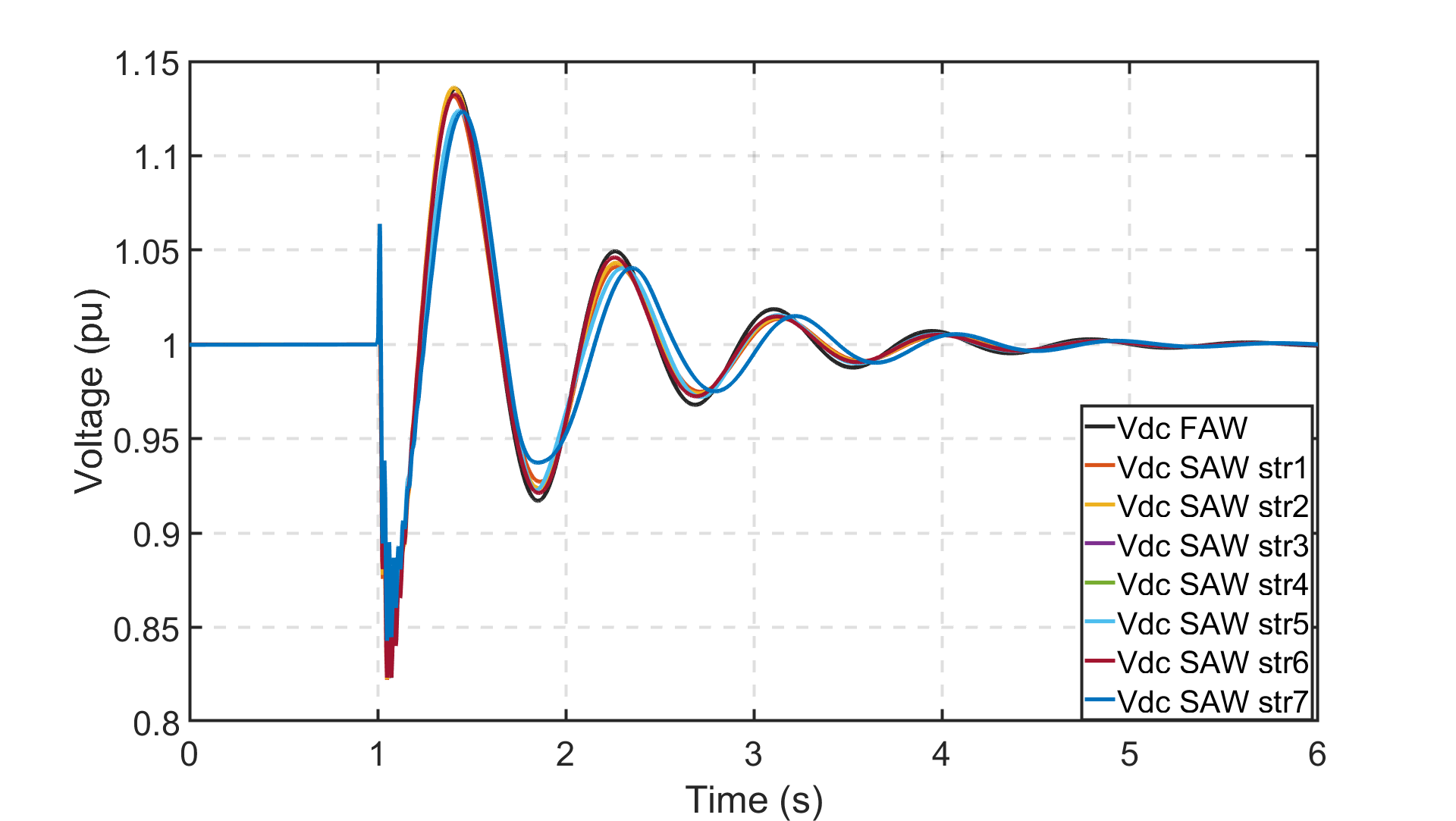}
    \caption{DC link voltage of the FAW, SAW with unequal generation, at grid voltage phase shift of of $15^{\circ}$}
    \label{fig:vdc}
\end{figure}

However, when the power generation of the WTG's in the WF was different, even though the total power output remained the same as FAW, the difference in the dynamics between the full aggregation and string aggregation was apparent. This aspect is demonstrated in Fig. \ref{fig:phase shift 3}, where FAW and SAW with equal generation among strings and unequal generation among the strings are subjected to a $15^{\circ}$ grid phase jump. The damping ratio has been intentionally reduced to demonstrate the difference in dynamics in aggregation methods.
The dc-link voltage for the same event is shown in Fig. \ref{fig:vdc}.
The dc link voltages is different between the strings during the phase shift event which cannot be captured by FAW. The differences in dc-link voltage dynamics for a system event can also lead to the loss of information in FAW modeling. It should be noted that disturbances levels as shown in Fig. \ref{fig:vdc}, in practise could be enough to trigger the dc chopper within the converter, which, as discussed earlier, has not been included in this study.

The key conclusion is that a FAW model is only adequate in capturing the dynamics if all the strings are symmetrical with simillar electrical and control and generation levels between the strings in the SAW. The divergence between the FAW and the complete model of the WF, including all the WTG, could be even more profound.

\subsection{Adequacy of the aggregated modelling With Current limit}
The GFC- WTG is equipped with a current limit which limits the current at 1.2 pu of the rated. This paper studies the impact of the current limit in a GFC-WF and is restricted to current limit triggering during the phase jump event which happens nearly instantaneously. 

\begin{figure}
    \centering
    \includegraphics[width=\linewidth]{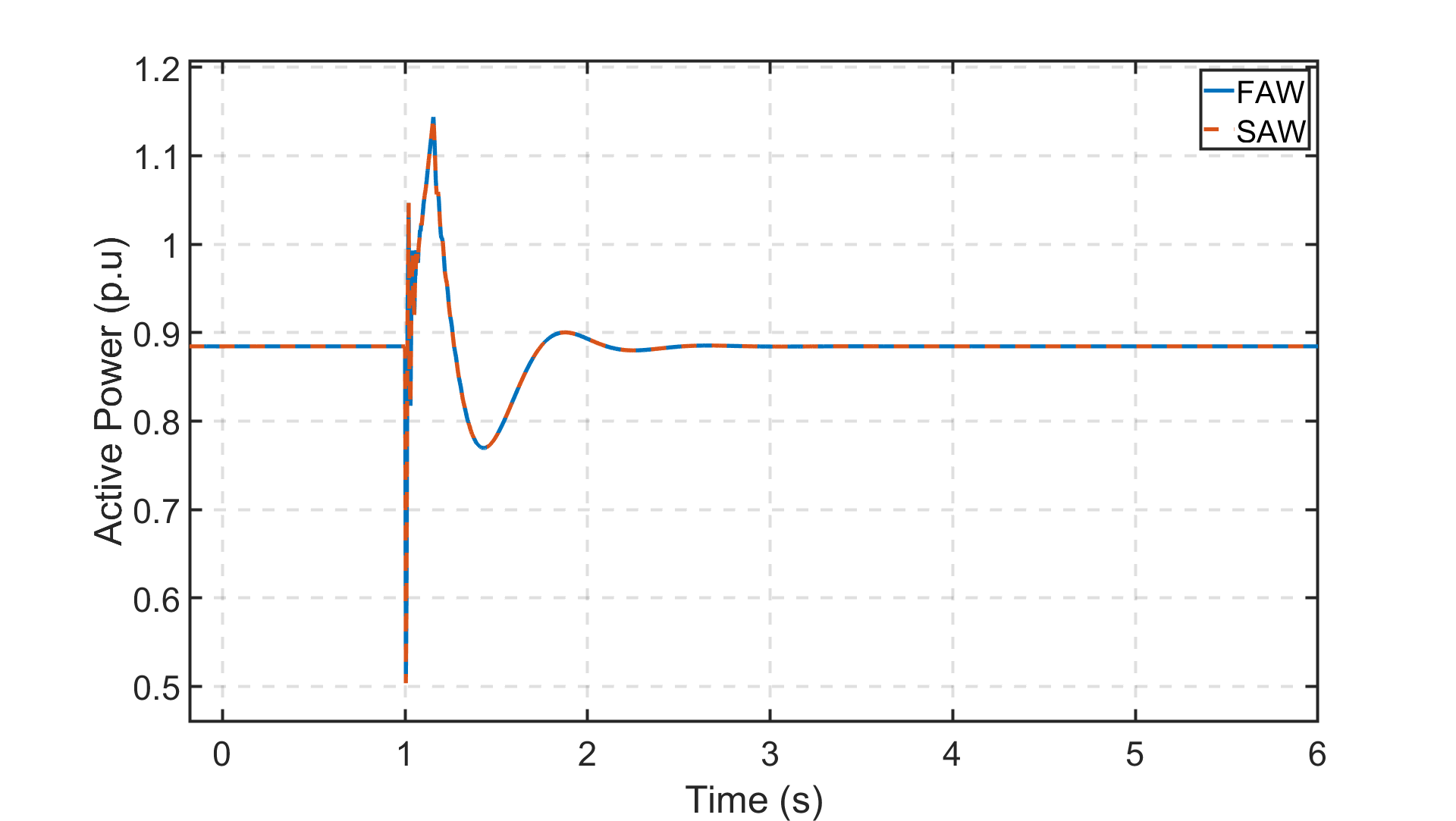}
    \caption{Output power of both SAW and FAW under a $20^{\circ}$ grid phase jump event with equal equal 0.9 pu power generation with current limit at 1.2 pu}
    \label{fig:phae shift with cl}
\end{figure}

The output power of both SAW and FAW under a $20^{\circ}$ grid phase jump event with equal 0.9 pu power generation with a current limit at 1.2 pu is shown in Fig. \ref{fig:phae shift with cl}. Due to the symmetricity among the WTG strings, the dynamic response between both the models is similar.

\begin{figure}
    \centering
    \includegraphics[width=\linewidth]{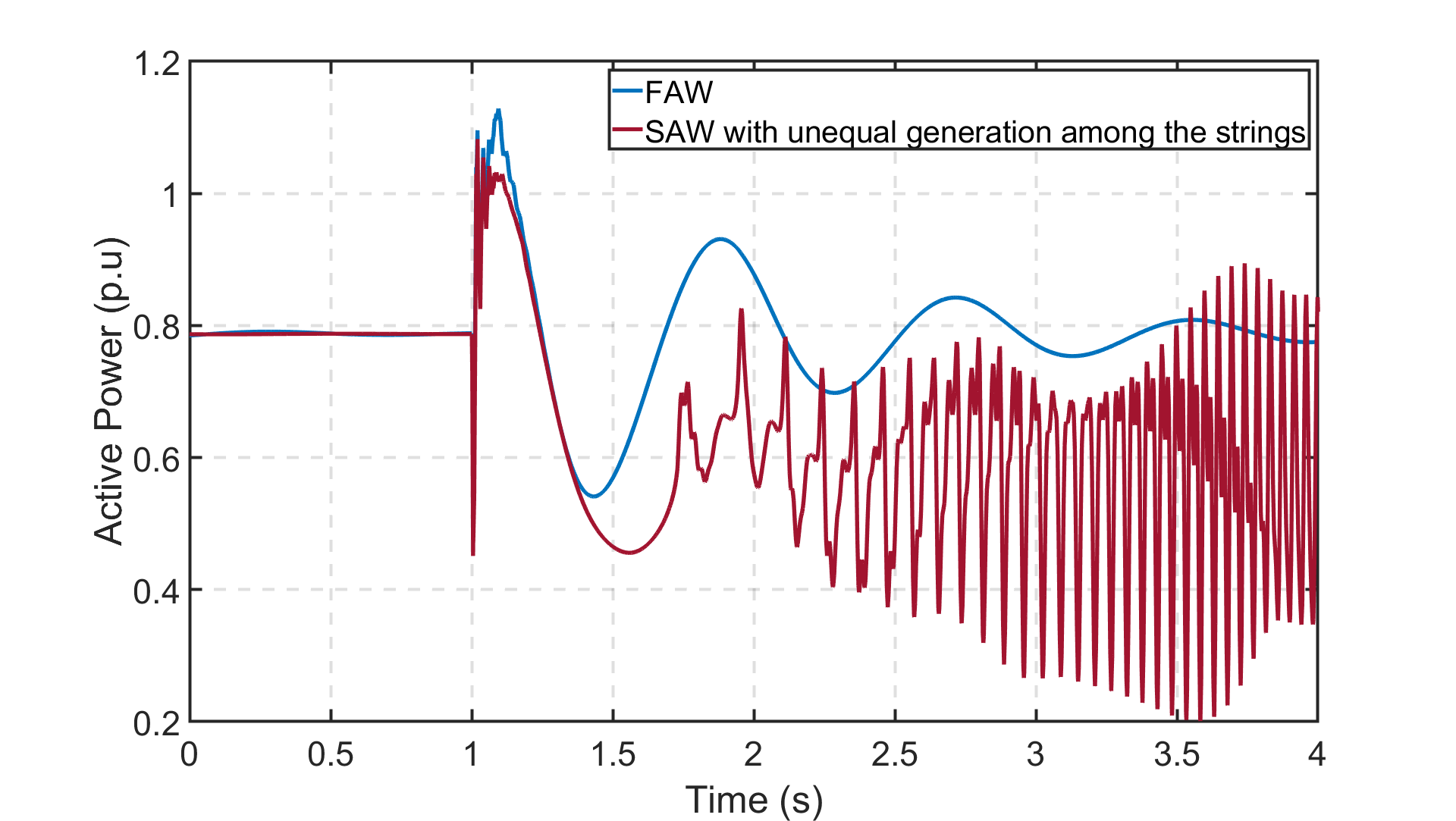}
    \caption{Output power of both SAW (unequal generation) and FAW under a $20^{\circ}$ grid phase jump event with 0.8 pu power generation with current limit at 1.2 pu}
    \label{fig:phae shift with cl2}
\end{figure}

Evaluating the results from FAW shown in Fig. \ref{fig:phae shift with cl}, one might conclude that the GFC-WF can survive a grid voltage phase shift of $20^{\circ}$ at an overall power generation of 0.9 pu. However, the individual loading of the WTGs is more critical here. For instance, Fig. \ref{fig:phae shift with cl2} shows the output power of both SAW (but unequal generation among strings) and FAW under a $20^{\circ}$ grid phase jump event with 0.9 pu power total generation. It is seen that there is a large disturbance in the output of the SAW-WTG, resembling the loss of synchronism dynamics of a synchronous generator.

\begin{figure}
    \centering
    \includegraphics[width=\linewidth]{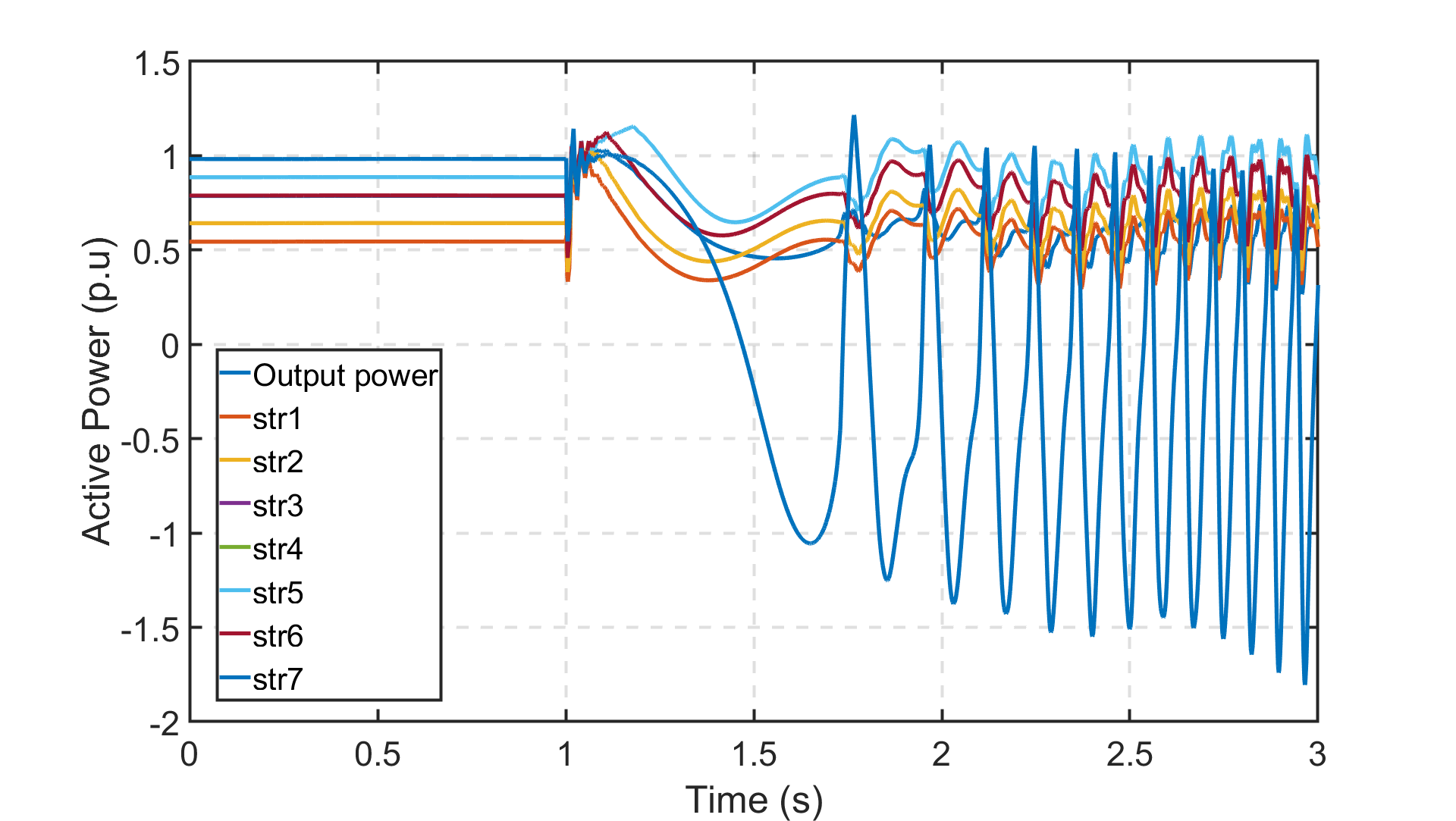}
    \caption{power output of of individual strings of SAW, with net generation of 0.8 pu power unequally distributed among the strings with limit at 1.2 pu}
    \label{fig:phae shift with cl3}
\end{figure}

The power output from each of the seven strings (str1-str7) from SAW for the same event is shown in Fig. \ref{fig:phae shift with cl3}, the power output of the strings are depicted in its own base power (60 MVA). It can be seen that the string with pre-event power generation close to 1 pu loses synchronism first and corrupts the power output from the rest of the WTG strings. 
Also, subsequent simulations analysis confirmed that the stability margin for GFC-WF against phase jump events needs to be assessed with the WTG/string with the largest generation modeled and not for the fully aggregated WF.

\subsection{Impact of heterogeneous parameters emulated inertia and damping parameters among the strings}

\begin{figure}
    \centering
    \includegraphics[width=\linewidth]{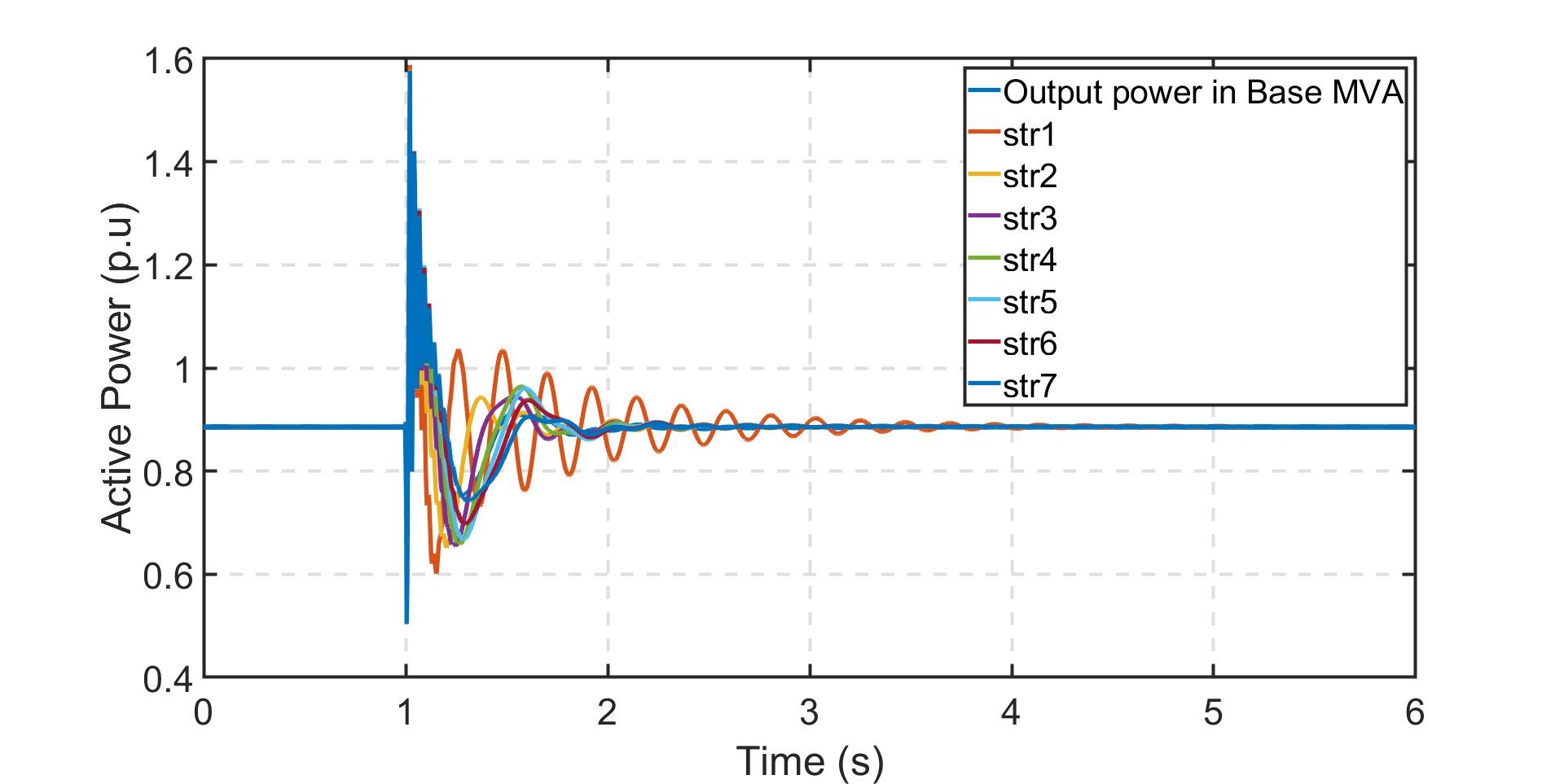}
    \caption{Power output of of individual strings of SAW, with equal generation of 0.8 pu but different inertial and damping parameters, without current limit}
    \label{fig:phae shift with cl4}
\end{figure}

\begin{figure}
    \centering
    \includegraphics[width=\linewidth]{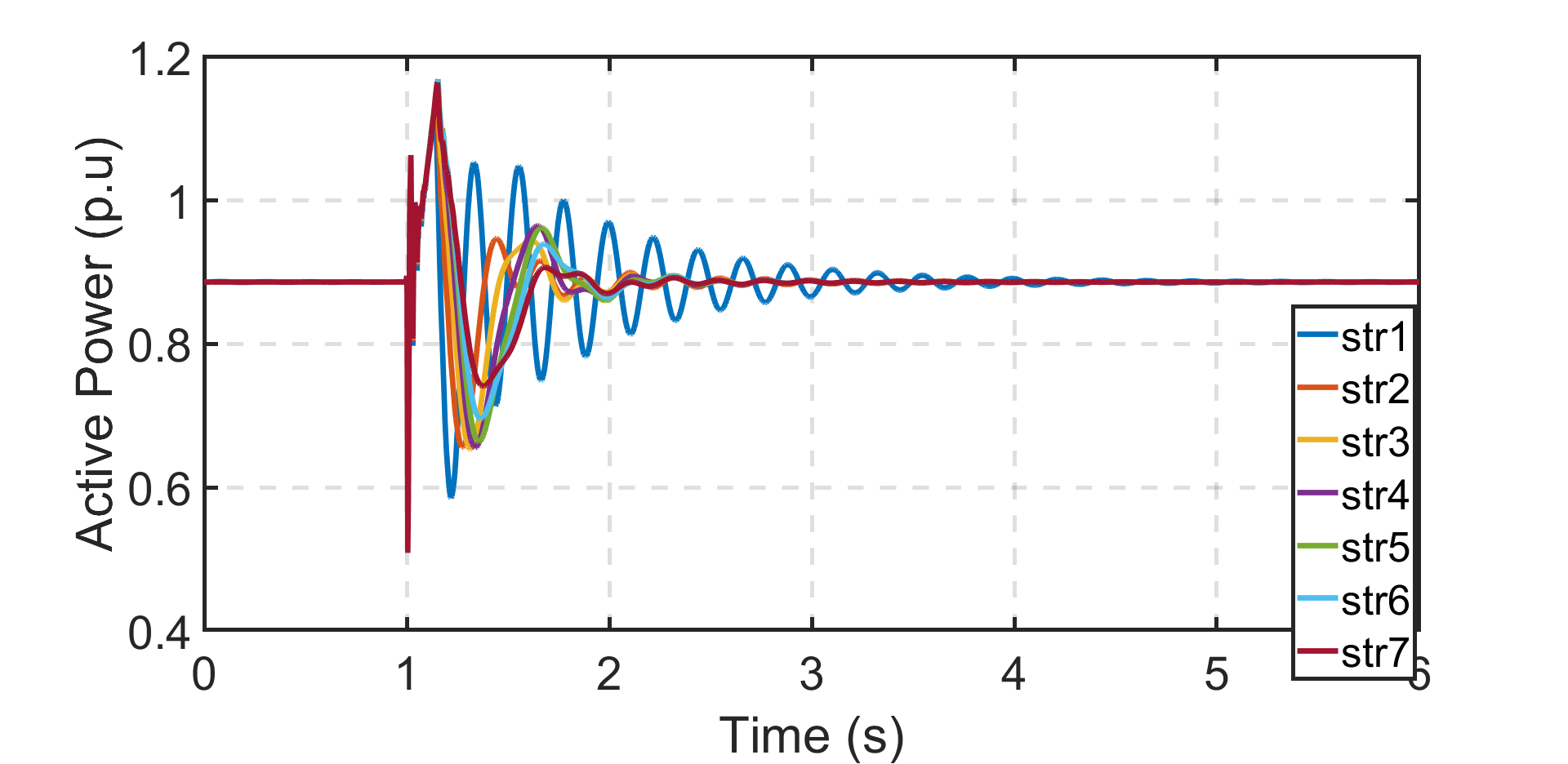}
    \caption{Power output of of individual strings of SAW, with equal generation of 0.8 pu but different inertial and damping parameters, with current limit}
    \label{fig:phae shift with cl5}
\end{figure}

In the previous subsections, all the strings in the SAW had the same inertia and damping constant. In this subsection, analysis is conducted with all the seven-strings in the string level aggregated WTG operating with different inertial and damping ratios. The objective is to evaluate any adverse interaction between the GFC-WTG if programmed with different inertia and damping characteristics. The GFC without the current limit is considered first.  The damping ratio of the strings is varied from [0.5 0.8], with the first string (str1) having the lowest inertia and damping and the last string having the highest inertia and damping (str2). Also, the power output of the strings are depicted in its base power (60 MVA). The power outputs from the strings for a phase jump event (15 degrees) without and with current limit is shown in Fig. \ref{fig:phae shift with cl4} and \ref{fig:phae shift with cl5}. No adverse interaction among the GFC-WTGs were observed due to unequal inertial and damping parameters.

\section{Discussion and Conclusion}

In this paper, the EMT model of a fully aggregated grid forming wind farm represented using a single WTG and EMT model of the grid forming wind farm aggregated at the string level is developed. An assessment of the single aggregation adequacy revealed that when the WTGs has diverse power generation levels, the aggregated WF represented by a single WTG fails to capture the full dynamics. Also, diverse power generation among the WTGs in a WF is common because of the wake effect and spatial distribution of the wind farm. The studies on GFC-WTG with the current limit demonstrated that a single fully aggregated model incorrectly captures the stability margin for a potential loss of synchronism due to phase jump event. The simulation studies confirmed that the stability margin for GFC-WF against phase jump events needs to be assessed with the WTG/string with the largest generation modeled and not on the fully aggregated WF. During the model development, it was found that the current control bandwidth had to be slowed down for GFCWTG to operate stably without triggering any instability. However, slowing down the GFC current control bandwidth can contradict GFC requirements. Therefore, further studies on control design to meet small signal stability and large signal requirements need to be conducted to further optimize the control and maximize the GFC-WTG performance.


\section*{Acknowledgment}
The study is funded by Phoenix project, funded by Ofgem under Network Innovation Competition programme, Project Direction ref: SPT / Phoenix / 16 December 2016 (https://www.spenergynetworks.co.uk/pages/phoenix.aspx).



%





\vspace{-1.0cm}	

\end{document}